# Does it matter which search engine is used? A user study using post-task relevance judgments

Sebastian Sünkler
Hamburg University of Applied Sciences
Finkenau 35, 22081 Hamburg, Germany
sebastian.suenkler@haw-hamburg.de

Dirk Lewandowski
Hamburg University of Applied Sciences
Finkenau 35, 22081 Hamburg, Germany
dirk.lewandowski@haw-hamburg.de

**ABSTRACT**
The objective of this research was to find out how the two search engines Google and Bing perform when users work freely on pre-defined tasks, and judge the relevance of the results immediately after finishing their search session. In a user study, 64 participants conducted two search tasks each, and then judged the results on the following: (1) The quality of the results they selected in their search sessions, (2) The quality of the results they were presented with in their search sessions (but which they did not click on), (3) The quality of the results from the competing search engine for their queries (which they did not see in their search session). We found that users heavily relied on Google, that Google produced more relevant results than Bing, that users were well able to select relevant results from the results lists, and that users judged the relevance of results lower when they regarded a task as difficult and did not find the correct information.

## INTRODUCTION
Search engines are an important means for finding information on the Web, and because of the Web's importance to knowledge acquisition, the are also an important means to what users get to know online. Users predominantly use Google, especially in the European countries, where Google has a market share of well over 90% (comScore, 2013). This raises the – admittedly not new – question whether Google is really better than its competitors in providing the user with relevant results. While the question is old, we still lack methods for realistically and reliably comparing search engines in a natural setting. In this paper, we propose a method and present a user study aiming at allowing for such comparisons.

The objective of our research was to find out how the two search engines Google and Bing perform when users work freely on pre-defined tasks (i.e., formulating their own queries and determining the length of their search sessions), and judge the relevance of the results immediately after finishing their search session. Approaches taken so far either focus on comparing the results of search engines using jurors who are not aware of where each result comes from (for an overview, see Lewandowski, 2015), on comparing search engines' results ranking to users' rankings of the same results (Bar-Ilan, Keenoy, Yaari, & Levene, 2007), or on measuring task success or user satisfaction in user studies where users use either the one or the other search engine (cf. White, 2016, p. 328ff.).

In this paper, we describe a method and an empirical study extending user-centered approaches by adding system-oriented tests to the model, and also by developing software support for efficiently conducting such studies. We conducted a lab-based user study with 64 participants. All search queries used in the users' sessions were sent to Google and Bing to collect the top ten results to the queries. Participants were given all of the collected results and asked to judge their relevance. So, users were able to judge (1) the relevance of the results for their own queries, (2) the results for all queries within their session, and (3) results coming from a search engine they had not used.

## LITERATURE REVIEW
### Users' satisfaction with search engine results
When users are asked, the majority states that they are satisfied with the results quality of search engines (Purcell, Brenner, & Raine, 2012), and "91% of search engine users say they always or most of the time find the information they are seeking when they use search engines" (Purcell et al., 2012, p. 3). A search engine's ranking of results is even considered as a criterion for credibility (Westerwick, 2013). Furthermore, users usually do not reflect on the relevance calculations made by search engine algorithms and the resulting results ordering (Tremel, 2010). Users most often choose only from the first results page, and they prefer the first few results listed (Joachims et al., 2007). Petrescu (2014) reports that more than two-thirds of all clicks go to the first five positions, and the result ranked first alone accounts for 31% of all clicks. In a large-scale study analyzing millions of queries from the Yahoo search engine, Goel at al. (2010) found that only 10,000 different websites account for approx. 80% of clicks on the search engine results pages (SERPs). Users are generally satisfied with the first few results, even when the results positions are mixed, and therefore, less relevant results are shown on the first position(s) (Keane, O'Brien, & Smyth, 2008; Pan et al., 2007). Users generally prefer Google's results to those from other search engines. This may have to do with quality issues, but studies also found that branding plays an important role (Jansen, Zhang, & Schultz, 2009). This raises the question whether the perceived superiority of Google can be confirmed when comparing this search engine's results to the ones from its competitors. There is a vast body of research on comparing commercial search engine results (for overviews, see Lewandowski, 2008, 2015). Most recently, Lewandowski (2015), in a large-scale retrieval effectiveness study, found that while Google outperforms Bing on navigational queries, Google's results for informational queries are, on average, only slightly more relevant than Bing's. He conjectured that a user would not recognize these small differences when using these two search engines. Schaer, Mayr, Sünkler, & Lewandowski (2016) compared top-ranked results with so-called "long tail" results, i.e., results shown on lower results positions. They found that the top results are judged as only slightly more relevant than the long tail results and concluded that the long tail provides a rich resource, as it provides the user with *different*, although still relevant results.

## Measuring retrieval effectiveness

Retrieval effectiveness studies rely on the design of "classic" Cranfield-style information retrieval tests and in cases of investigating web search engines adjust these methods (Gordon & Pathak, 1999; Griesbaum, 2004; Hawking, Craswell, Bailey, & Griffiths, 2001). In general, such tests consist of the following steps: A sample of search queries will be sent to the information systems under investigation, then the returned results will be collected, their ranking position randomized and the source anonymized to avoid learning and branding effects. After that, jurors judge the relevance of the results. Then, the results will be allocated again to the search engines and analyzed by using established evaluation metrics like recall and precision. This approach is sometimes considered as being too narrow, as it does not focus on the users' side of the search. However, conducting retrieval effectiveness studies in such a controlled environment still has its merits (Voorhees, 2009), although most researchers agree that these Cranfield-style tests should be replenished by user-focused studies. While Cranfield-style studies are usually conducted to compare the relevance of the results from different information retrieval systems, in the context of web search, the approach has also been used for questions going beyond that. For instance, researchers considered the relevance of sponsored results versus organic results (Jansen, 2007) and the commercial intent of results providers (Lewandowski, 2011). Information retrieval evaluation has seen a turn from system-centered towards user-centered evaluation (Kelly, 2009). It has become clear that while Cranfield-style studies (still) have their benefits, focusing on the user perspective (and process-oriented metrics, see White (2016), p. 309ff.) adds significantly to improving information retrieval systems. We also argue that focusing not only on the results to a single query but all the results seen by a user in a search session would greatly add to our understanding of concepts such as search engine bias and the influence of commercial search engines on what information users actually consume. User-centered studies are often criticized for the low number of users investigated, for the choice of participants (often undergraduate students), and for their lack of control of effects that search engines' branding and interfaces might have. From this short review of the two paradigms of information retrieval evaluation, it becomes clear that there is a need for combining the two approaches, or at least taking elements from the two approaches to building new evaluation frameworks combining "the best of the two worlds".

## FRAMEWORK TO EXTEND USER-BASED STUDIES

For our approach, we first analyzed given frameworks and models (Belkin, Cole, & Liu, 2009; Borlund, 2003). Based on that analysis we aimed to create a simple and flexible framework to extend user tests by explicit relevance judgments. Our aim was a model where users judge results they have actually seen in the test, as well as other search results, shortly after they worked on a task. Overall, our method is separated into nine parts, some of which are decisions researches have to make when designing a test (cf. Gordon & Pathak, 1999; Hawking et al., 2001; Tague-Sutcliffe, 1992), and therefore, these parts are independent of any software implementation: (1) Selection of search engines to investigate. To collect natural user interactions, users should be able to select a search engine of their choice to solve tasks. It should be possible to define a set of search engines (not necessarily used by the users when working on their tasks) for relevance judgments, to support a comparison between search systems. (2) Definition of user groups. (3) Definition of test timeframe and test environment (making it possible to conduct long-time studies in natural environments or lab studies). (4) Design of questionnaires (allowing users to, e.g., rate difficulty, outcome, effort, and learning success of tasks). (5) Design of simulated search tasks. (6) Definition of scales and questions about the search results (e.g., binary decisions, Likert-scales). (7) Collecting interaction data, taking into account the search engine(s) used, results clicked, and the search queries used. (8) Build result sets based on search engines used, search queries, clicked results and additional search results from other search engines, which have not necessarily been used by the actual user in the study. (9) Analyze user interactions and results judgments.

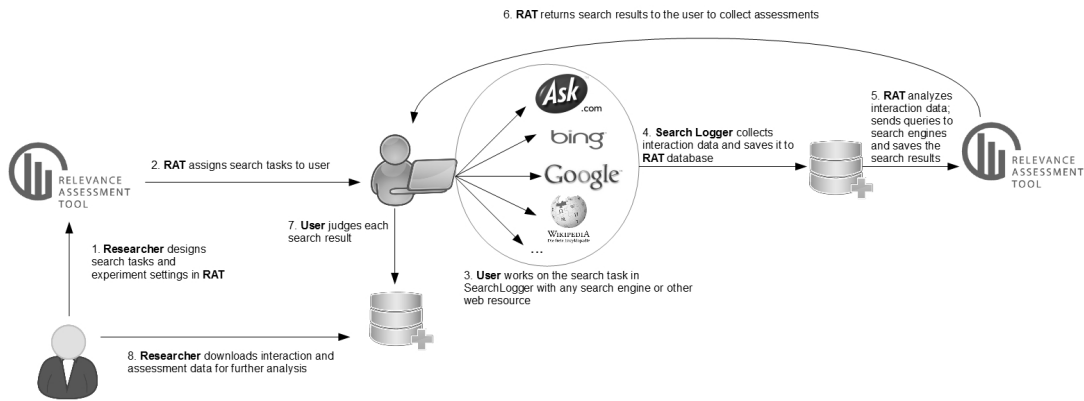

**Figure 1. Process of the model in practice**

In theory, all of these parts can be used in studies without using special software. Parts 1 to 6 are independent of any software, as these are the parts where researchers have to make these decisions for the experimental set-up. Parts 7 and 8, however, cannot be practically conducted without software support: Researchers would need to collect all logging data to create the results sets they want to have judged by users. For example, they would need to send the search queries to the search engines themselves and save the results locally. This would make it impossible to let participants judge the results shortly after they finished the task.

Fig. 1 shows the simple process and the resulting possibilities when our framework is implemented in software. Users work on search tasks using a search engine or other web resource of their choice while all interactions are logged in the background. When they finish the task, the logged data get analyzed. These data contain timestamps stating when a user started and finished a task, all search queries used, clicks on search results and the web pages opened. The software also has an option to gather data using questionnaires before starting a task, and after finishing it. This makes it is possible to add demographic data and statements to the tasks or anything else researchers may want to know from their users. After analyzing the log data, search queries are extracted from the log data and sent to the search engines under investigation. The returned search results are saved through screen scraping (up to a defined maximum of results, e.g., the top ten results for each query). All other websites used by the participant will be added to the returned results to create the result set the user has to judge later. To avoid bias or branding effects, the user will not see the source of the given website she has to assess (i.e., the name of the search engine that produced the result).

In summary, this new approach of combining a task-based user study with a system-oriented retrieval effectiveness study has the following advantages: (1) The participant can formulate her own queries, and therefore, realistic search queries are used. (2) Relevance judgments are made by the same person that formulated the queries. This is based on the assumption that the person who formulated the query statement is best able to judge on the results to that query. (3) We can directly compare the results from the search engine a participant actually used to the results from any competing search engine.

## SOFTWARE DEVELOPMENT

We developed software to use our method in practice. As said in the introduction, one obstacle to conducting user-centered retrieval effectiveness studies is the lack of software tools supporting such studies. There are some tools available to evaluate search systems (for an overview, see Sünkler, 2013), but they are limited in different ways. Most of them lack supporting tests with web search engines or they just allow for using their Application Programming Interfaces (APIs). However, search results from APIs differ from the results seen by a search engine user (Mayr & Tosques, 2005). Furthermore, only a few commercial web search engines provide an API. Most importantly, Google does not. Further disadvantages of existing tools are that they either only focus on a specific use case (Tawileh, Griesbaum, & Mandl, 2010) or were even developed for and used in a single project only (Fox, Karnawat, Mydland, Dumais, & White, 2005; Machill, Neuberger, Schweiger, & Wirth, 2004; Pan et al., 2007). More sustainable tools already available either focus on rather fragmented tasks (like the Digital Methods Tools), on crawling for web data analysis (like SocSciBot and Webometric Analyst), or on study designs based on test collections (like the Lemur Toolkit and Revelation)[1]. Our aim, therefore, was to build software to use with search engines that are actually used by Internet users on an everyday basis. The Relevance Assessment Tool (RAT) is a software toolkit that allows researchers to conduct large-scale studies based on results from (commercial) search engines and other information retrieval systems (Lewandowski & Sünkler, 2013)[2]. It consists of modules for (1) designing studies, (2) collecting data from search systems, (3) collecting judgments on the results, (4) downloading/analyzing the results. A starting point to developing RAT was the fact that retrieval effectiveness studies usually require a lot of manual work in designing the test, collecting search results, finding jurors and collecting their assessments, and in analyzing the test results, as well. RAT addresses all these issues and aims to offer help in making retrieval studies more efficient and effective. While RAT allows for efficiently conducting retrieval effectiveness tests, it does not support any user interactions with the search engines to be analyzed. The Search Logger (Singer, Norbisrath, Vainikko, Kikkas, & Lewandowski, 2011) is a tool developed to log user interactions in a Web browser. While there are several such tools available (Capra, 2010; Jansen, 2006), the Search Logger was our choice, as it provides options to design search tasks and logging features to collect interaction data related to the designed search tasks but also came from developers with whom we had already worked. The Search Logger allows for designing search tasks as well as pre- and post-task questionnaires. It is a Firefox Browser Add-on that offers a simple interface for letting users start search tasks. Interaction data are collected in the background as a user uses search engines and browses websites. Actions from that data which are important for our software are: user starts a new task, user finishes a task, used search engines in a task, used search queries in a task, opened tabs, clicked links (for details on the data the Search Logger can collect, see Singer et al., 2011, p. 753). While the search logger allows for collecting interaction data conveniently, its output is only raw data. Therefore, software based on the Search Logger first has to analyze the log output and extract data from it for further investigation. The basic idea of the software designed for extending system-orientated tests by user interactions is described in our framework above. For this purpose, we built an application that analyses the Search Logger output, extracts queries and search engines used from it, and feeds the resulting data into the Relevance Assessment Tool.

---

[1] See https://wiki.digitalmethods.net/Dmi/ToolDatabase, http://socscibot.wlv.ac.uk, http://lexiurl.wlv.ac.uk.

## RESEARCH QUESTIONS

As said in the introduction, the objective of our study was to find out how the two search engines Google and Bing perform when users work freely on pre-defined tasks and judge the relevance of the results immediately afterwards. This resulted in the following research questions:

RQ1: To what extent do users use different search engines when working on search tasks given by the experimenter?

RQ2: Do users judge results they actually selected when performing their tasks as more relevant than results not selected during the task?

RQ3: Would users be equally satisfied with the results of another search engine; i.e., would Google users be equally satisfied with results from Bing?

RQ4: Would users judge search results as more relevant if they themselves judged the retrieved information as correct to solve the task?

RQ5: Do the two search engines perform differently when considering simple vs. complex tasks?

While the first four research questions relate to users' behavior when searching for information, the last two research questions relate to comparing the results quality of the two dominant commercial search engines.

## METHODS

### Choice of search engines

We selected the two search engines Bing and Google, as these are the two major commercial search engines. Given the predominant use of Google in Germany (the country where we conducted our research), it is interesting to consider its major, though concerning usage far behind, competitor Bing. Does this search engine produce results at least comparable to Google's?

### Tasks and participants

We recruited a convenience sample of 64 participants. We aimed at creating a sample that consisted of adolescents and adults of all ages and an even gender distribution. The average age was 37.2 years (SD=12.29). 31 participants were female and 33 male. With such a sample, we hope to get better insight into actual searching behavior, even though we may face problems with statistical significance (Singer, Norbisrath, & Lewandowski, 2012). Each user was given two tasks, one simple and one complex. Each user worked on different tasks, and a total 60 different search tasks were used, evenly divided between simple and complex. For creating the search tasks, we used tasks from a previous study (Singer, Norbisrath, et al., 2012), as well as tasks we derived from questionnaires given to students in our department. Following Singer, Norbisrath, & Lewandowski (2013), we define simple tasks as lookup tasks that can be satisfied with just one document. Examples we used included searches for birthdates and birthplaces of scientists, news facts, historical facts, facts about television and movies, and geographical information. Complex tasks are tasks in which a user has to collect information about a topic from multiple documents and has to use more than one query to successfully complete the task. Examples we used included the collection of information for leisure activities, essential information for a specific purchase, and different recreational opportunities at a certain location. For a more detailed discussion on simple and complex tasks, as used in this study, see Singer, Danilov, & Norbisrath (2012). We also used pre- and post-task questionnaires. We showed the participants the task and asked them to judge its difficulty, and whether they felt able to find the correct information related to the stated information need, as well. After they had finished the task, we asked them the same questions again, this time relating to their actual performance.

---

[2] While the software toolkit is not available for download, we invite researchers to contact us if they are interested in using RAT for their research.

**Data Collection**
Data were collected in our lab in Hamburg, Germany. While users were allowed as much time as they needed to complete their tasks, we planned 30 minutes for each participant, which turned out to be more than sufficient. While users were allowed to use any search engine (or other search tool) they knew, all participants solely relied on the Google search engine. Participants first addressed the simple tasks and then judged the relevance of the results shown in that session before completing the complex tasks. Again, they judged the results after completing the task. Users were allowed to use their own queries. This leads to a more realistic setting, even though we are aware that the wording of the task descriptions may have influenced the way participants formulated queries. The users acted in interactive query sessions. They were able to refine their search queries to adjust the given search results to solve the tasks. We also asked users for assessing the difficulty of the given tasks and whether they found the correct information to solve the task. The corresponding questions were, "The search task was easy to solve (yes/no)", and, "I found the correct information (yes/no)", respectively. For their simple task, seven participants said it was not easy to solve it, and ten participants said that for their complex task. Overall, seven participants thought they did not find the correct information (three simple tasks, four complex tasks).

*Collecting session data*
Participants started their search sessions by opening the Search Logger plugin in Firefox. First, they saw the simple task and answered the pre-task questionnaire for it. The participants were asked to assess the difficulty level, to estimate the time for the task and whether they would be able to find the correct information to solve the search task. Then, they were free to choose whatever tool(s) they wanted, search engine or not, to solve the task. They were allowed to spend as much time as they liked on the task, and when they were finished, they had to fill in the post-task questionnaire. They were again asked about the difficulty level, their actual time effort, the number of search queries, and whether they thought they found the correct information to solve the task. Session data collected by the Search Logger was then automatically loaded into the Relevance Assessment Tool. Processing the data there took a few minutes; therefore, participants had a short break before judging the results. After judging the results, the participants opened the Search Logger again to work on the complex task. The process was similar to the work on simple tasks. The users answered the same pre-task questionnaire, used a search engine or other tool(s) of their choice, and then answered the post-task questionnaire. Then, they judged the collected search results from their search session.

*Collecting relevance judgments*
From the logged session data, the following were extracted: Search queries, search engine used, and results selected on the SERPs. The Relevance Assessment Tool then automatically queried the search engine the participants had used (which was Google in all cases) and collected the first ten results for each query. The same queries were sent to Bing, as well, and the results were collected accordingly. For practical reasons, we had to limit the maximum number of queries sent to each search engine to three. We assumed that it would not be reasonable to give participants more than 60 results (3 queries × 10 results × 2 search engines) to judge. The collected documents were presented to the participants in a random order. Results from both Google and Bing were mixed, and participants did not know the source of a particular result or whether they had selected that result during the first part of the study. Duplicates (i.e., URLs returned by both search engines) were filtered out, and a participant only had to judge them once. Participants were then asked to judge the relevance of each result. We asked users to judge each result first on a binary scale (relevant or not) and then on a 5-point Likert scale[3].

Due to limitations caused by actual user behavior during the test and technical difficulties we were not able to use all collected session data to create the result sets for judgments. Some participants simply used the descriptions on the SERPs to solve tasks; others relied on Wikipedia to find the required information, and therefore, did not use a search engine at all. In total, participants assessed 1,156 results from Google and 1,132 from Bing, from a total of 101 search tasks.

**RESULTS**
**Click distribution on SERPs**
First, we looked at the distribution of clicks on the Google SERPs (Fig. 2). The graph shows that participants heavily relied on the results order presented by the search engine. These findings are in agreement with previous studies (e.g., Goel et al., 2010; Pan et al., 2007). However, we have to keep in mind that in seven cases, participants only used the information on the SERPs or used other web resources like Wikipedia to solve the task, and therefore, these tasks did not result in any clicks on the SERPs.

**Results precision (all results)**
One aim of our study was to compare the participants' preferred search engine (Google) with its major competitor, Bing. To answer RQ3, we investigated the differences in users' satisfaction with the retrieved results. We plotted

---
[3] Labels of the scale were: 1 = completely irrelevant, 2 = irrelevant, 3 = relevant, 4 = highly relevant, 5 = completely relevant. We used the same scale as in Lewandowski (2015).

precision graphs for all results from both search engines (Fig. 3).

We can see that Google results are, on average, judged better on all results positions. However, the differences between Google and Bing are small. This finding is again in agreement with previous studies (e.g., Lewandowski, 2015). Calculating standard precision measures for measuring retrieval performance does not show significant differences in retrieval effectiveness of the two search engines in most cases, except for MAP@5 (see Table 1).

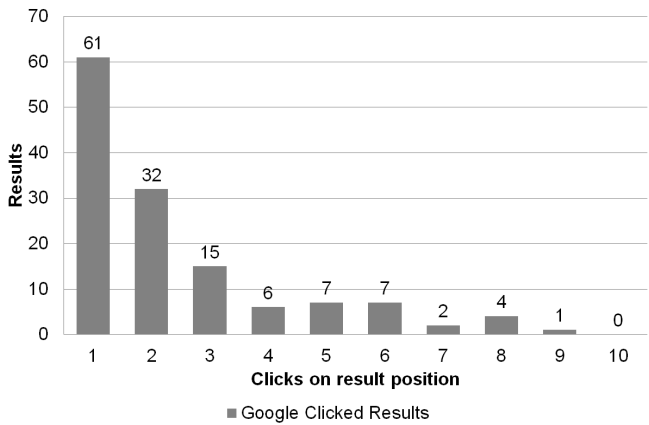

**Figure 2. Distribution of clicked results on result positions in Google**

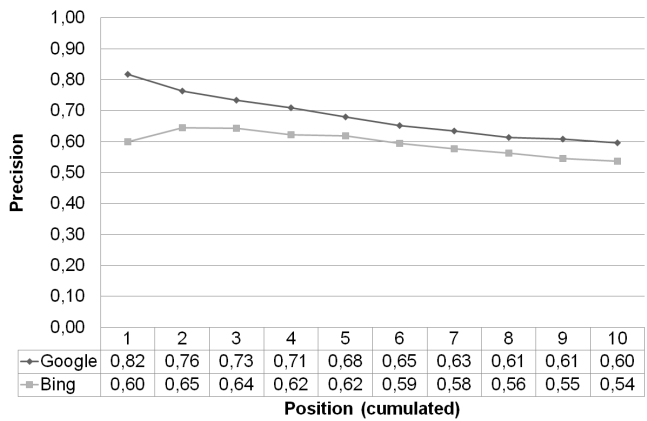

**Figure 3. Precision graph for Google and Bing results**

When looking at the more differentiated scale judgments (Fig. 4), we can see that Google outperforms Bing. Especially noteworthy are the results judged as "completely irrelevant," which account for approximately 36% of all Bing results.

**Results precision (clicked results)**

We measured the precision of *clicked results* to investigate RQ2 and to find out if users judge selected results as being more relevant than results not selected during the tasks. As all users chose Google for their tasks, we only have click data for that search engine. The precision graph in Fig. 5 shows that regardless of results position, the precision of the clicked results lies at approx. 0.8, which is higher than the precision for all results, even those positioned first (see Fig. 3). This means that users are well able to judge the relevance of the results presented when selecting results based on descriptions (snippets) on the SERPs. The ratio of relevant results to relevant snippets is in agreement with results from a study by Lewandowski (2008) that compared users' judgments of the relevance of results descriptions with their judgments of the actual results.

**Table 1. Retrieval results for Google and Bing using a two-tailed Student's t-test (α <= 0.05) results on time P@5 (p=0.256), P@10 (p=0.259), MAP@5 (p=0.035), MAP@10 (p=0.098), NDCG@5 (p=0.244) and NDCG@10 (p=0.181)**

|  | P@5 | P@10 | MAP@5 | MAP@10 | NDCG@5 | NDCG@10 |
|---|---|---|---|---|---|---|
| Google | 0.68 | 0.60 | 0.85 | 0.79 | 0.81 | 0.90 |
| Bing | 0.62 | 0.54 | 0.74 | 0.72 | 0.78 | 0.88 |

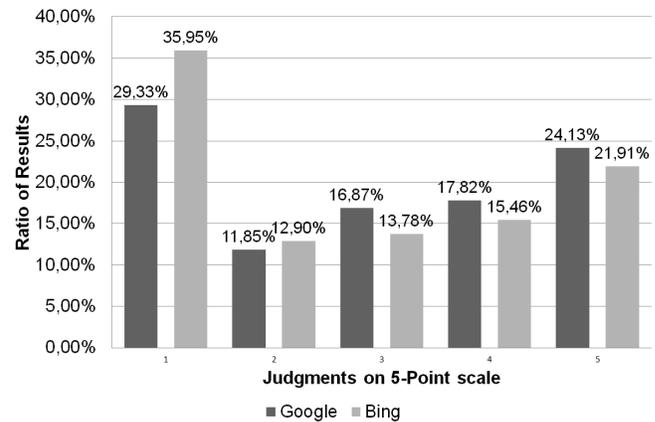

**Figure 4. Distribution of judgments on the 5-point scale**

Looking at the distribution of scale judgments (Fig. 6), we can see that users are well able to select results that they later (when they see the result instead of the snippet only) judge as highly relevant. 36% of the results selected are judged as completely relevant, and taken together, results judged as highly relevant or completely relevant account for approximately 58% of all selected results.

When looking at the total number of clicks on each result position and their relevance as judged on the 5-point scale, we can see that results for the first positions are not only clicked more often but are also considered more relevant. It should be noted that even though participants were not able to see the results position, remembering documents seen during the first part of the study might have had an influence on the judgments.

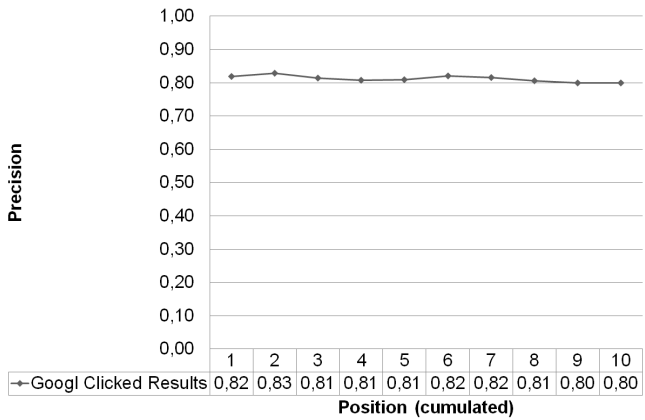

**Figure 5. Precision graph for clicked results (Google)**

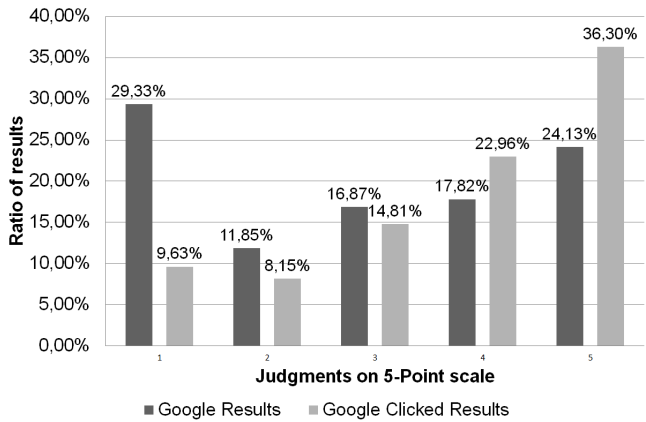

**Figure 6. Distribution of judgments for clicked results**

Taken together, we can see that the relevance ranking (as expressed through results position and users' selections) and selection decisions based on results descriptions go hand in hand and lead to users getting relevant results to their searches.

**Results precision (task difficulty, task success)**

Another aim of our study was to investigate how relevance judgments correlated with task difficulty and task success. We asked participants to assess the difficulty of a task and their success on it after they had finished it. First of all, we calculated the average precision for all search results for Google and Bing using the binary judgments. We found an average precision of 0.57 for Google and 0.50 for Bing. Most participants said they had found the correct information and judged a task as not being difficult. Even if we see differences in precision scores between both search engines, and also in comparison to the general average precision of both search engines, we cannot state any significant correlation between task difficulty and relevance judgments or task success and relevance judgments due to the low amount of assessed search results.

We also analyzed the judgments on the 5-point-scale. We found no significant differences between successful tasks and unsuccessful tasks. The same applies to 5-point scale judgments for difficult and not difficult tasks, except of results rated as completely irrelevant (1 point on the scale). In Google, participants rated 30% of the results as completely irrelevant, compared to 36% on Bing. Compared to all given judgments on the scale (see Fig. 4), participants tended to judge results lower when they rated a task as difficult.

**Overlap between search engines**

To calculate the overlap of search results from Google and Bing we followed the method used by Spink and Jansen (Spink, Jansen, Blakely, & Koshman, 2006). First, we removed all query duplicates. This resulted in 119 search queries from 50 simple and 51 complex tasks, leading to 1,888 unduplicated search results. 1,545 results (81.8%) were unique to one search engine, 343 (18.2%) were shared by both search engines. 777 results were unique to Bing, 768 results unique to Google. This finding shows that even when the two search engines are comparable regarding relevance, a user may still benefit from using the other search engine to get different results.

**Simple vs. complex search tasks**

Next, we investigate the influence of task complexity (RQ5). Table 3 shows the number of clicked results on the SERPs for simple and complex tasks. The data show that participants clicked on significantly more results for the complex tasks. However, the number of results selected is quite low. In some cases, it was possible to find an answer already on the SERP, so no results selection was needed. This explains the minimum numbers in Table 3.

**Table 3. Statistics on tasks complexity using a two-tailed Student's t-test (α <= 0.05) on time effort (p=0.114), search queries (p=0.757) and clicked results (p=0.119).**

|  |  | N | Min | Max | Mean | Std. Deviation |
|---|---|---|---|---|---|---|
| Time effort | Simple task | 64 | 6 | 1085 | 157.13 | 154.73 |
|  | Complex task | 64 | 37 | 774 | 195.13 | 112.45 |
| Search queries | Simple task | 64 | 0 | 8 | 1.53 | 1.31 |
|  | Complex task | 64 | 1 | 5 | 1.59 | 0.94 |
| Clicked results | Simple task | 64 | 0 | 5 | 1.48 | 0.96 |
|  | Complex task | 64 | 0 | 7 | 1.81 | 1.37 |

Details about average time needed to complete the tasks and the average number of search queries used are also shown in Table 3. Surprisingly, there are no significant differences between the number of queries needed for the simple and the complex tasks. However, there is a significant difference between the median time needed to complete these two task types. While other studies found task time and queries per task to be good measures for distinguishing between simple and complex tasks (Singer, Norbisrath, et al., 2012), our data may be explained by user characteristics. Furthermore, since our users worked first on the simple tasks and then on the complex ones, it may well be that they put more effort into the first task and therefore spent more time and queries on them.

## DISCUSSION

As users often solely rely on a single search engine, user studies comparing different search engines are flawed because users are aware of the search engines that are used. Therefore, branding effects may play a larger role than the actual differences in results quality. Furthermore, when users are accustomed to a certain user interface, they may face problems with other interfaces and therefore judge the quality of these search engines as being lower. The method presented in this paper addresses this problem by letting users choose the search engine of their preference and then comparing datasets from that search engine to the results from another one.

In our study, users solely relied on using Google. Although users were free to use any search tool they liked, none of our 64 participants decided to use any alternative (RQ1). However, we can only speculate on the reasons for this, the most obvious one being that users are so used to Google that they did not think of any alternatives. This is also illustrated by Google's overwhelming market share in Germany.

Regarding RQ2, we see that results actually selected are judged better than the results not selected. From this, we can see that users are well able to make their decisions about relevant results based on results positions and descriptions, meaning that the snippets are helpful and users can use them to their benefit. This leads us to question the exclusive use of precision-based measures in search engine results evaluations. It may even be questionable to use the whole results set in retrieval effectiveness studies. While it is surely important to know how many results are irrelevant when a larger number of results needs to be examined (at least on the SERPs), some irrelevant results may not be too problematic when Web searchers tend to select only one or a few results that just need to be "good enough."

Regarding RQ3, we found that the results from Google were judged as being more relevant than those from Bing, but the differences are not too big (and for most measures applied not significant). This is in line with findings from previous studies (Lewandowski, 2015) and suggests that it is not superior quality that explains users' heavy reliance on Google, but other factors probably play a role, as well. This leads to the conclusion that while Bing also produces a large ratio of relevant results, there is no motivation for users to switch to Bing as their standard search engine. This would only be the case if Bing produced results of an appreciable better quality. However, users may have good reasons for using Bing as an *additional* search engine. This is illustrated by the low overlap between the results from the two search engines.

Regarding RQ4 and RQ5, we tried to investigate if task success and task difficulty have influences on the relevance judgments of the participants in the study. We found differences in binary judgments, as well as in graded judgments on the 5-point scale. Users tended to judge results lower when they were not able to find the correct answer and when they rated a search task as difficult. However, we cannot assume a statistically significant correlation between task success and relevance judgments or task difficulty and relevance judgments because, probably due to the low amount of collected search results.

## CONCLUSION

In this study, 64 participants each worked on two tasks, using whatever search tool they liked, and formulating their own queries. Their log data was automatically analyzed immediately after task completion, and results from the search engine used for solving the tasks (which in all cases was Google) and from a competing search engine (Bing) were automatically collected. All collected results were given to the same user who conducted the search for judging their relevance. With this approach, we were able to compare the results quality of the two search engines, controlling for brand effects and interface issues.

Maybe the most important finding of this study is that while participants judged Google's results as slightly more relevant than Bing's, the differences are not too big. Only considering results quality, we can assume that users would also be satisfied with using Bing, as we assume that they would not recognize the small differences. Furthermore, we found that the overlap between the results from the two search engines is low. Therefore, users would benefit from using Bing (or other search engines) as an addition to Google, as they would get a different picture to their information need.

Our research has several limitations. Results could have been influenced by choice of participants, although we took great care of having a varied sample regarding age and gender. More severely, our choice of tasks might have affected the results, i.e., the search engines could have performed differently when using different tasks (cf. Lewandowski, 2008). However, to avoid this effect, we used a large variety of tasks, distributed randomly over users.

On the software side, the contribution of this paper is a new framework for interactive information retrieval evaluation, focusing on Web search, combining a task-based user study with a system-oriented retrieval effectiveness study. Our research offers a framework for studying actual users' behavior in correlation with relevance judgments more deeply than before. The major advantage to this approach is that it allows studies with users formulating their own queries, going through whole search sessions, and then judging the relevance of their own results. Furthermore, we can use the Relevance Assessment Tool to collect results from search engines that users did not use and inject these results into the set of documents the participants are given to judge.

While our research was designed as a lab study, the software could also be used in any setting where participants use a Firefox browser. All the user needs to do is select the "Firefox to go" installation option provided by the tool.

In future research, we plan to conduct further studies using the framework presented in this paper. A limitation of our study was that while our new approach was presented, we were not able to systematically compare it with conventional retrieval effectiveness studies. A first step for future work would be to set up two such studies and then compare whether our approach leads to more meaningful results. Another step would be conducting studies with bigger samples to investigate correlations between user-based metrics and relevance judgments. Conceivably, such studies could be accomplished as long-term studies in "natural" environments.